\documentclass[preprint2,tighten,times]{aastex}  
\pdfoutput=1

\usepackage{
	amsmath,
	amssymb,
	amsthm,
	booktabs,
	comment,
	epsfig,
	graphicx,
	grffile,
	import,
	morefloats,
	microtype,
	natbib,
	subfigure,
	times,
	ulem,
	url,
	verbatim,
	wasysym,
	xspace
}

\newcommand{\as}{\mbox{\arcsec}}

\newcommand{\kms}{\mbox{\ km s$^{-1}$}} 

\def\lsim {$\rlap{\raise.4ex\hbox{$<$}}\lower.55ex\hbox{$\sim$}\,$}
\def\gsim {$\rlap{\raise.4ex\hbox{$>$}}\lower.55ex\hbox{$\sim$}\,$}

\newcommand{\hillf}{\textsc{hill{\footnotesize 5}}}

\newcommand{\hcop}{${\rm HCO}^+$} 

\input{epsf}

\begin{document}

\title{
Searching for Inflow Towards Massive Starless Clump Candidates Identified in the Bolocam Galactic Plane Survey
}
\shorttitle{Searching for inflow}
\shortauthors{Calahan et al.}

\author{
    Jenny K. Calahan$^{1,2}$,
    Yancy L. Shirley$^{1}$,
    Brian E. Svoboda$^{1}$
    Elizabeth A. Ivanov$^{1}$
    Jonathan R. Schmid$^{1}$
    Anna Pulley$^{1}$
    Jennifer Lautenbach$^{1}$
    Nicole Zawadzki$^{1}$
    Christopher Bullivant$^{1}$
    Claire W. Cook$^{1}$
    Laurin Gray$^{1}$
    Andrew Henrici$^{1}$
    Massimo Pascale$^{1}$
    Carter Bosse$^{1}$
    Quadry Chance$^{1}$
    Sarah Choi$^{1}$
    Marina Dunn$^{1}$
    Ramon Jaime-Frias$^{1}$
    Ian Kearsley$^{1}$
    Joseph Kelledy$^{1}$
    Collin Lewin$^{1}$
    Qasim Mahmood$^{1}$
    Scott McKinley$^{1}$
    Adriana M. Mitchell$^{1}$
    Daniel R. Robinson$^{1}$
}
\affil{
	$^1$Steward Observatory,
	University of Arizona,
	933 North Cherry Avenue,
	Tucson,
	AZ 85721,
	USA;
	\href{mailto:jcalahan@email.arizona.edu}{jcalahan@email.arizona.edu} \\
	$^2$Senior Honors Thesis submitted at The University of Arizona\\
	\vspace{-6mm}
}

\begin{abstract}
Recent Galactic plane surveys of dust continuum emission at long wavelengths have identified a population of dense, massive clumps with no evidence for on-going star formation.  These massive starless clump candidates are excellent sites to search for the initial phases of massive star formation before the feedback from massive star formation effects the clump.  In this study, we search for the spectroscopic signature of inflowing  gas toward starless clumps, some of which are massive enough to form a massive star. We observed 101 starless clump candidates identified in the Bolocam Galactic Plane Survey (BGPS) in HCO$^+$ J = 1-0 using the 12m Arizona Radio Observatory telescope. We find a small blue excess of \(E=(N_{blue}-N_{red})/N_{total}=0.03\) for the complete survey. We identified 6 clumps that are good candidates for inflow motion and used a radiative transfer model to calculate mass inflow rates that range from 500 - 2000 M$_{\odot}$/Myr.  If the observed line profiles are indeed due to large-scale inflow motions, then these clumps will typically double their mass on a free fall time.
Our survey finds that massive BGPS starless clump candidates with inflow signatures in HCO$^+$ J = 1-0 are rare throughout our Galaxy.
\end{abstract}

\maketitle


\section{Introduction}\label{sec:Introduction}
Massive (M $> 8$ M$_{\odot}$) stars form from dense prestellar cores within massive molecular clumps \citep{McKee07}.  The theories of the dominant physical mechanisms responsible for their formation are still polemical \citep{Motte17}. Do massive protostars form from the monolithic collapse of a massive (M $> 30$ M$_{\odot}$) prestellar core or do protostellar cores accrete surrounding material and grow rapidly through a competitive accretion with the surrounding environment?  This long standing problem in the field of star formation can be addressed by looking for signature of large scale accretion, determining how often it occurs, and the rate at which material is accreting.  In this paper, we present a spectroscopic survey of the optically thick, intermediate density gas tracer, HCO$^+$ 1-0, searching for the signature of large scale inflow in massive clumps in the Milky Way for which no signature of star formation has been previously detected.

Recent blind surveys of the Galactic plane at far-infrared through submillimeter wavelengths \citep[i.e.\ BGPS, ATLASGAL, Hi-GAL, JPS;][]{Ginsburg13, Urquhart14, Elia17, Eden17} have identified a population of tens of thousands of massive clumps (M $> 200$ M$_{\odot}$) which do not appear to be forming stars.  These objects were classified as Starless Clump Candidates (SCCs) in the evolutionary analysis of clumps identified in the 1.1 mm Bolocam Galactic Plane Survey \citep{Svoboda16}.  The typical SCC has a median mass of $230$ M$_{\odot}$, a size of 1 pc, a gas kinetic temperature (measured from NH$_3$ observations) of $13$ K, and a starless phase lifetime that scales inversely with the mass of the clump \citep{Svoboda16}.  The clumps are starless candidates because current Galactic plane surveys are limited to detecting protostars with luminosities L $>\sim 30$ L$_{\odot}$ at a distance of a few kpc \citep{Svoboda16}.  Even if there is a population of undetected low-mass protostars already forming in them, SCCs represent environments that are in a early stage of star formation and which have not been severely disrupted by feedback from intermediate or high-mass protostars \citep{Matzner17}.  SCCs are ideal locations to search for evidence of large scale flows that may be important in the formation of massive protostars.

A spectroscopic signature of inflow is the presence of a blue asymmetric, self-absorbed line profile meaning that the line peaks at smaller velocities with respect to the $v_{LSR}$ of the object and the line shape is double-peaked with a clear self-abroption dip.  The general conditions for the formation of this line profile are that the tracer must be optically thick enough to create self absorption with increasing excitation temperature and a favorable velocity field along the line of sight (i.e. the flow have a component projected along the line of sight).  The presence of a blue asymmetric self-absorbed profile is not evidence of inflow alone as other physical mechanics (namely rotation and outflow) can also mimic similar line shapes under the appropriate conditions (see \cite{Evans03}).  Generally, a convincing case can be made when the spectral line profile is mapped over a core and radiative transfer models are used with independent constraints on the source physical structure, for instance a density profile derived from dust continuum observations, to successfully model the observed line profile.  Convincing collapse signatures seen in emission have been observed toward individual dense cores \citep[e.g.][]{Choi95,Myers95,Narayanan98,Tafalla98,Evans05,Seo11}.  Recently there have been observations of purported large-scale flows toward high-mass \citep{Peretto13}, intermediate mass \citep{Kirk13}, and low-mass \citep{Palmeirim13}  star-forming regions.  A survey searching for the signature of inflow motions has not been performed yet toward the newly identified SCCs in the BGPS.
A survey for inflow motions towards SCCs has the advantage that their kinematics should not be dominated by outflows since there is no detected star formation and therefore a blue asymmetric, self-absorbed profile is more likely to be associated with inflow.

\begin{figure}
\centering
\includegraphics[scale=.39,trim={0 50mm 0 15mm},clip]{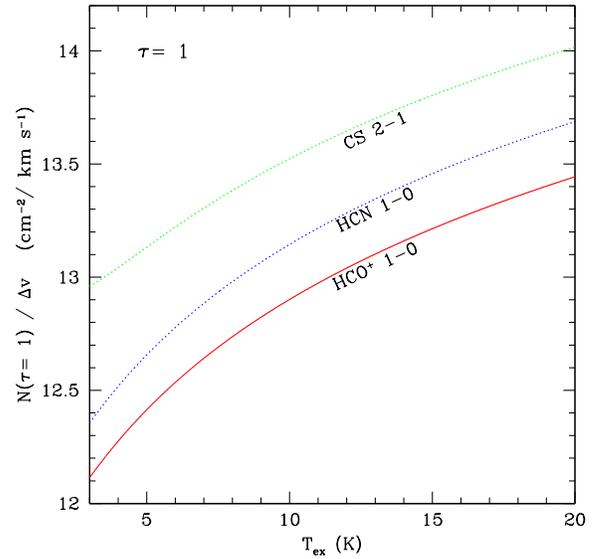}
\caption{The column density at which a molecular tracer becomes optically thick (defined here as $\tau = 1$) at the line peak plotted as a function of the excitation temperature.  The curves were calculated assuming a FWHM linewidth of $\Delta v = 1$ km/s. The HCN 1-0 curve has not accounted for hyperfine structure; the curve for the strongest HCN 1-0 hyperfine transition (F = 2-1) would be the plotted curve multiplied by 9/5 = 1.8.}
\label{opthick}
\end{figure}

For this survey, we chose the HCO$^+$ 1-0 line as an optically thick kinematic tracer of intermediate density gas in SCCs.  HCO$^+$ 1-0 has an effective excitation density of $\sim 1000$ cm$^{-3}$ \citep{Shirley2015}.  We can assess the column density at which a particular molecular tracer becomes optically thick using
\begin{equation}
\frac{N_{\tau = 1}(T_\mathrm{ex})}{\Delta v} = 4 \sqrt{\frac{\pi^3}{\ln{2}}} \frac{\nu_{ul}^3}{g_u A_{ul} c^3} \frac{Q(T_\mathrm{ex})}{e^{-E_l/kT_\mathrm{ex}} - e^{-E_u/kT_\mathrm{ex}}}
\end{equation}
where $T_{ex}$ is the excitation temperature of the rotational transition from $u \rightarrow l$ with energy levels $E_u$ and $E_l$, frequency $\nu_{ul}$, statistical weight $g_u$ and Einstein $A_{ul}$.  $Q(T_{ex})$ is the partition function assuming every energy level has the same $T_{ex}$ (the CTEX approximation; see \cite{Mangum15}).  The threshold optically thick column density $N_{\tau = 1}$ is a function of $T_{ex}$ and the FWHM linewidth ($\Delta v$ given in  cm/s) where we have assumed a Gaussian line profile.  Figure \ref{opthick} shows this threshold column densities for commonly observed dense gas tracers assuming a FWHM linewidth of 1 km/s.  Typical HCO$^+$ column densities observed toward clumps in Galactic surveys are in the range $10^{13} - 10^{15}$ cm$^{-2}$ \citep{Shirley13,Hoq13} which are well above the threshold column density curve for all excitation temperatures $T_{ex} < 12$ K. HCO$^+$ 1-0 is an excellent optically thick tracer of intermediate density, sub-thermally populated gas with which to search for the kinematic signatures of inflow.
Simulations of global collapse agree with this assertion and further illustrate that the 1-0 transition is the best transition with which to search for inflow signatures \citep{Smith2013}.

In this paper, we present a systematic survey of 101 SCCs located toward the first quadrant
of the Milky Way in HCO$^+$ 1-0 using the 12m Arizona Radio Observatory telescope.
We present the observations (\S2) and analyze the line profiles searching the signature of inflow
(\S3).  We find 6 inflow candidates and model their line profiles using the HILL5 radiative transfer model (\S4).  Finally, we calculate mass inflow rates and discuss how this may affect the evolution of these clumps (\S4). 

\section{Observations}\label{sec:ObservationsSelection}

\begin{figure}
\centering
\includegraphics[scale=.39]{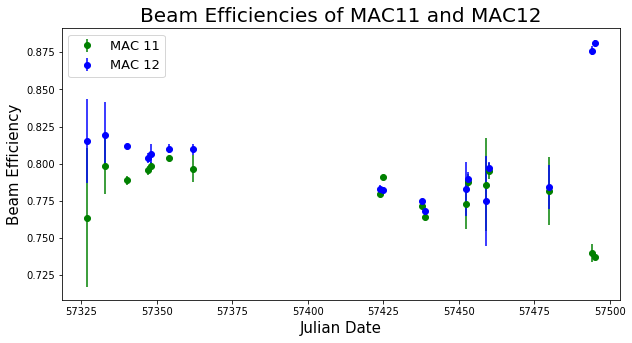}
\caption{Main beam efficiencies derived each day for each polarization (MAC 11 and MAC 12) and used for that day's data reduction. For MAC 11 the median value is 0.787 and for MAC 12 it is 0.801. }
\label{beameff}
\end{figure}

We observed at the Arizona Radio Observatory (ARO) 12m radio telescope on Kitt Peak for 19 observing shifts between September 2015 and April 2016 \footnote{This project was the group radio observing project of the undergraduate Astronomy Club at The University of Arizona}. We tuned to the J = 1-0 transition of HCO$^+$ at 89188.5250 MHz in the lower sideband using the ALMA Band 3 ARO prototype dual polarization sideband-seperating reciever. Each polarization of the lower sideband was connected to the Millimeter AutoCorrelator (MAC) spectrometer (denoted MAC11 and MAC12) with 0.02 km/s (6.1 kHz) resolution. The FWHM beamsize at this frequency was $67.6$\as .  Each source was observed for 60 minutes of total integration time where we position switched between our source and an OFF position every 30 seconds. We identified clean OFF positions (no HCO$^+$ J=1-0 emission at the velocities of our sources) along the Galactic Plane every half-degree. Observations of  Jupiter and Venus were used to calibrate the spectra from the T$_A^*$ scale to T$_{mb}$ scale during each shift \citep{Mangum93}. Figure \ref{beameff} shows the beam efficiencies for each polarization. The median \(\eta_{mb}= 0.787\) for MAC11 and \(\eta_{mb}=0.801\) for MAC12. Our average baseline RMS was $\sigma_{T_{mb}} = 0.067$ K for the complete survey. 

\begin{figure}
\centering
\includegraphics[scale=.7]{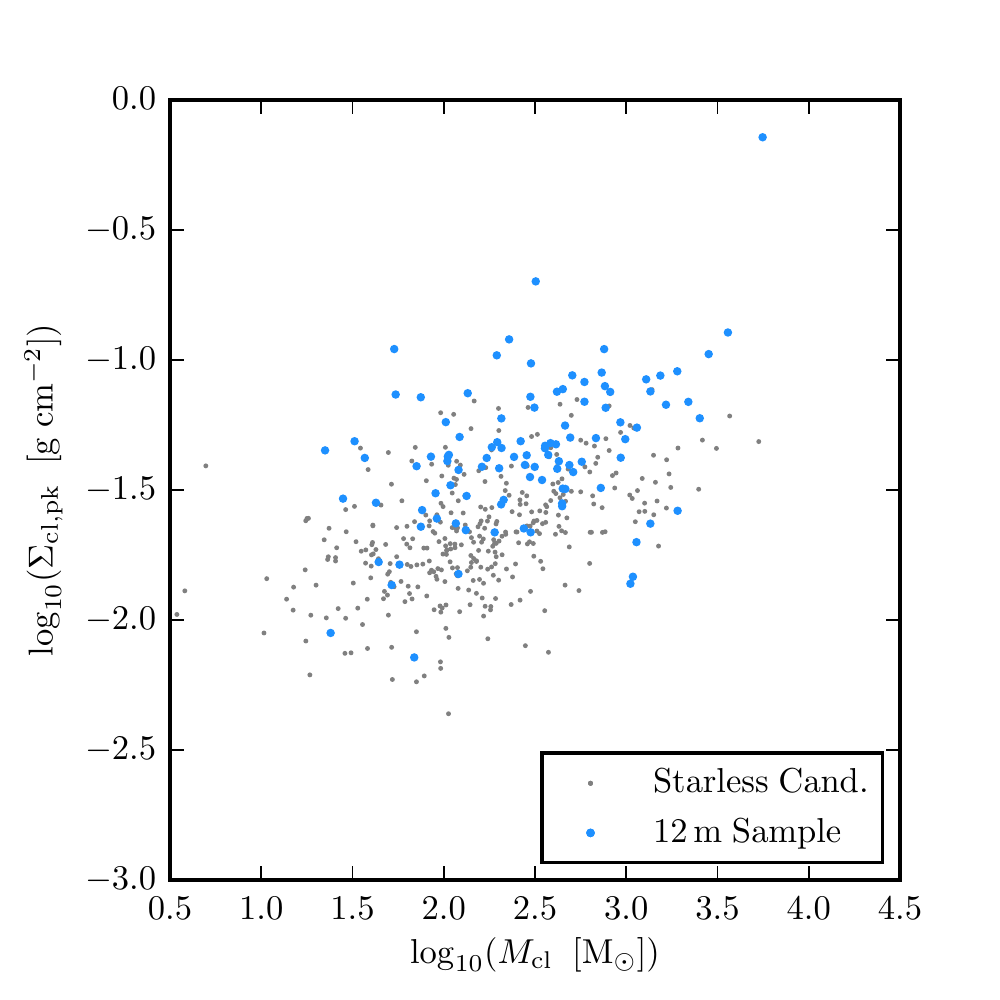}
\caption{The mass and peak mass surface density of BGPS starless clump candidates with well constrained distances. Each point corresponds to the median value of the mass and peak mass surface density probabilty density functions (see Svoboda et al. 2016).  "12m Sample" is the subset of sources observed in this paper. }
\label{sourceprop}
\end{figure}

Sources were selected blindly from the starless clump candidate list in \cite{Svoboda16} that had NH$_3$ (1,1) detections. The average distance of the sources was 4.3 kpc and range from 1.2-11.8 kpc. Our main beam size on the 12m is approximately 1.39 pc at that average distance. The average diameter of the sources is 2 pc, indicating that our single pointing observations cover most of the clump.  Many of the sources had also been previously detected in HCO$^+$ 3-2 \citep{Shirley13}, but since that survey sacrificed spectral resolution (only 1.1 km/s channel width) for wide bandwidth, we are unable to use the existing 3-2 observations to systematically search for inflow in the SCC sample.  Throughout this paper, we use the BGPS v2.0.1 catalog number \citep{Ginsburg13} for the source name (i.e. BGPS 4029).  The observed sample is 101 objects in the first quadrant of the Galactic Plane with a median distance of $4.1$ kpc (see Table 1). Figure \ref{sourceprop} shows the comparison of the mass and peak mass surface density between the 12m sample and the complete sample of SCCs with well constrained distances (see \cite{EBowers15} for an explanation of how distances are determined toward BGPS clumps).  The 12m sample has a median mass of $300$ M$_{\odot}$ and is representative of the full range of observed mass, spanning masses up to 5550 M$_{\odot}$ (BGPS 3114); however, the observed 12m sample is biased to higher peak mass surface densities by a factor of $\sim 2$ compared to the complete SCC sample with well-constrained distances (see \cite{Svoboda16}). 

For the best inflow candidates (\S3), we also made observations in H$^{13}$CO$^+$ 1-0 at 86754.2884 MHz.  The optically thinner H$^{13}$CO$^+$ observations were directly compared to the v$_{LSR}$ from NH$_3$ observations \citep{Svoboda16} as a secondary check that an optically thin line peaks in the HCO$^+$ 1-0 absorption dip.  The spectrometer setup and subsequent analysis are identical to that of HCO$^+$ 1-0 observations.

\section{Results}

\subsection{Line Asymmetry Statistics}

We observed a total of 101 starless clump candidates in HCO$^+$ J=1-0. Two of the spectra were not used for the subsequent analysis because they were non-detections at the NH$_3$ velocity (BGPS 3534 and BGPS 5183).  All spectra were visually analyzed to search for blue asymmetric line profiles with self-absorption dips.
If NH$_3$ peaked in the self-absorption dip of a blue asymmetric profile, it was considered a candidate for inflow. We found that six of the 101 sources (~6\% ) were good infall candidates. Their line profiles are shown in Figure \ref{candidates}. Spectra of all observed clumps are in the appendix. There are some negative velocities, which would suggest emission in the off positions, but any negative peaks are located far from the source spectra and  will not affect analysis. 

We checked these candidates by also observing the optically thinner tracer H$^{13}$CO$^+$ 1-0.  In all cases except for BGPS 2432, the H$^{13}$CO$^+$ peak velocity agreed with the NH$_3$ velocity within $0.2$ km/s.  Even in the case of BGPS 2432 where the velocity difference between H$^{13}CO^+$ and NH$_3$ was $+0.57$ km/s, both lines still peak within the HCO$^+$ absorption dip.  Thus H$^{13}$CO$^+$ observations confirm that an optically thinner line peaks in the absorption dip for all 6 candidates. 
This result also indicates that the NH$_3$ (1,1) and (2,2) transitions are good tracers of the systemic velocity of the clumps, partly due to their multiple hyperfine lines with different optical depths. 
In the subsequent analysis, we consider
the six sources (BGPS 2432, BGPS 3300, BGPS 3302, BGPS 3604, BGPS 4029, and BGPS 5021) as inflow candidates.

All of the six inflow candidates have an HCO$^+$ absorption dip that is slightly red of the NH$_3$ or H$^{13}$CO$^+$ peak velocities.  This effect has been observed in other inflow surveys \citep{He2016, Jin2016}.  As described in \citep{Jin2016}, global collapse (i.e. in competive accretion models) could potentially create a redshifted absorption dip.  This can be tested with higher resolution observations; at the single-dish resolution of observations in this paper, these sources remain only candidates for global inflow.

\begin{figure}
\centering
\includegraphics[scale=.44,trim={10mm 50mm 0 20mm},clip]{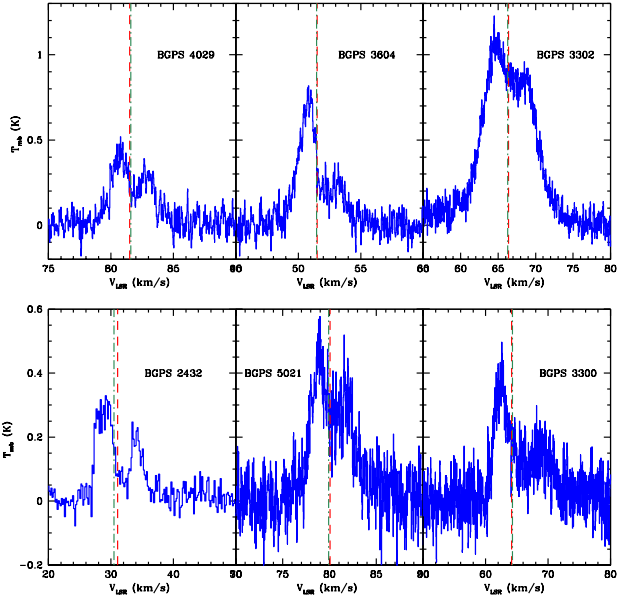}
\caption{Line profiles of our best inflow candidates. Each profile has blue asymmetry with an NH$_3$ peak (shown as the red dashed line) and H$^{13}$CO$^+$ peak (shown as a green dot and dashed line) in the self absorption dip. These were the only sources that met these two criteria  as well as had reasonable inflow velocities as modeled in \S4}
\label{candidates}
\end{figure}

\begin{figure*}
\centering
\includegraphics[scale=.6]{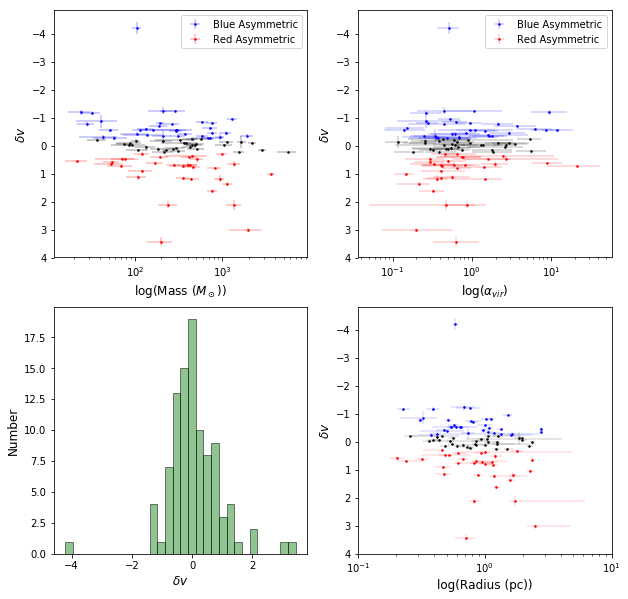}
\caption{
We compare \(\delta v\) to mass (upper left panel), virial parameter (upper right panel) , and clump radius (lower right panel). The histogram of \(\delta v\) values are shown in the  lower left panel. There is no obvious correlation present in the three \(\delta v\) comparison graphs. The spearman-rank coefficient of these three plots is 0.19 for mass, -0.12 for virial parameter, and 0.18 for radius. The red points are clumps with a red asymmetry \(\delta v> +.25\), and the blue points are clumps with a blue asymmetry \(\delta v < -.25\) while black points fall in a middle range of $\delta v$.}
\label{4panel}
\end{figure*}

A visual analysis of the spectra is not a robust quantitative measure of the asymmetry in the line profile; therefore we use \(\delta v\) to measure the velocity shift between HCO$^+$ and NH$_3$. \(\delta v\) was first defined by \cite{Mardones}:
\begin{equation}
\delta v = \frac{v_{HCO^+} - v_{NH_3}}{\Delta v_{NH_3}} \;\; ,
\end{equation}
where \(v_{HCO^+}\) is the peak HCO$^+$ velocity, \(v_{NH_3}\) is the peak NH$_3$ velocity, and $\Delta v_{NH_3}$ is the FWHM linewidth of the NH$_3$ lines. Ammonia parameters were taken from \cite{Svoboda16}. With this measure, if \( \delta v\) is negative, it has blue asymmetry. If it is positive, it has a red asymmetry.  

Figure \ref{4panel} show the histogram of $\delta v$ values for the observed SCC sample (also see Table 1).  Most of the sources are clustered within the range of $-2 \leq \delta v \leq 2$.  One notable exception is the source BGPS 3125 with a $\delta v < -4$.  Visual inspection of the source indicates that the HCO$^+$ 1-0 is a rare example of a triple-peaked spectrum with the NH$_3$ well aligned with a weaker peak in HCO$^+$.  Therefore, BGPS 3125 is not a good infall candidate and its large negative  $\delta v$ should be ignored.  
All of the 6 best infall candidates have $\delta v < -0.48$.  Of the remaining sources with $\delta v < -0.5$, eight of the sources are single-peaked in HCO$^+$ with no self-absorption dip, two sources are double peaked but the peaks are nearby equal in height (BGPS 2533 and BGPS 3151), one source is double-peaked but a poor inflow candidate due to NH$_3$ not peaking within the self-absorption dip (BGPS 2984), and three sources have more than 2 peaks in HCO$^+$. See Figures \ref{allspectra1} and \ref{allspectra2} in the Appendix for spectra of all observed clumps. 

There is a very slight bias of blue asymmetry over red asymmetry in our observed $\delta v$ distribution. We use the definition of blue excess given in \cite{Mardones}:
\begin{equation}
E=\frac{N_{\mathrm {blue}}-N_{\mathrm { red}}}{N_{\mathrm {total}}} \;\;,
\end{equation}
where \(N_{\mathrm {blue}}\) is the number of blue asymmetric profiles, \(N_{\mathrm {red}}\) is red asymmetric, and \(N_{\mathrm {total}}\) is the total number of clumps in our sample.
Using the limit that a \(\delta v\) is less than \(-.25\) is a significant blue asymmetry and greater than \(.25\) is a significant red asymmetry (since the $\pm 0.25$ threshold is approximately 5 times our median error in \(\delta v\)), we then have \(N_{blue} = \) 35 clumps and \(N_{red} =\) 32 clumps. This leaves another 32 clumps that had a \(\delta v\) value between \(-.25\) and \(.25\) (the total number of clumps with a valid \(\delta v\) value is 99 clumps).  Using these values we come to a blue excess of \(E=0.03\). 



Compared to other surveys that calculated the blue excess also using HCO$^+$, our value for blue excess is on the low end. \cite{He2015}, \cite{Fuller}, \cite{Wu}, \cite{Purcell06}, and \cite{Rygl} have each completed survey of various stages of star formation while using the chemical tracer HCO$^+$. Four out of the five of the surveys found excess values near \(E=.2\) while one found a value near ours at \(E=.02\). \cite{He2015} and \cite{He2016} used the MALT 90 survey data \citep{Jackson13} to calculate their blue excess, and their pre-stellar objects were most similar to our sample of SCCs; clumps with no mid-infrared sources identified by a flux cut in the ATLASGAL survey. We found a larger fraction of objects with red excess as compared to their results. 
We should note that there is significant difference between the definition of a inflow candidate between \cite{He2016} and our paper; namely, we require the line profile to have a blue asymmetry with a self-absorption profile while the \cite{He2016} paper only required that the source have $\delta v < -0.25$ in at least one optically thick tracer and no red skewed profiles in any optically thick tracer.  We find several sources with $\delta v < -0.5$ but which are single-peaked in HCO$^+$ 1-0.  With a single pointing survey and a large beam that encompasses the entire clump, we cannot confidently claim that these large negative $\delta v$ single-peaked HCO$^+$ 1-0 sources are good inflow candidates; the observed velocity offset betwen HCO$^+$ and NH$_3$ may be due to variations in the chemical or excitation structure of the clumps.
In addition to this, our sample was selected blindly from SCCs which had previously been detected in $NH_3$ emission while the MALT90 pre-stellar clumps were selected from a flux cut and therefore represent the brightest ATLASGAL clumps in that category.
 
The only study with a similar blue excess to ours is that of \cite{Purcell06} which observed methanol masers associated with massive YSOs. It is likely their E=0.02 is contaminated by strong outflow motions associated with massive protostars which can create a red asymmetric line profile. Our survey is unique in that we are observing the earliest stage of star formation where there should not be strong contamination due to outflows from intermerdiate or high-mass stars. With our value of \(E=0.03\) we see that the earliest stages of star formation does not strongly favor blue or red asymmetric HCO$^+$ 1-0 line profiles for BGPS starless clump candidates.

\subsection{Comparison with Physical Properties of the Clumps}

We compared \( \delta v\) to the virial parameter, and clump mass as seen in Figure \ref{4panel}. There is no strong relationship between \(\delta v\) and the mass, meaning that given a certain clump's mass we cannot accurately predict that same clump's \(\delta v\) value. We might have expected there to be a slight relationship that suggested more massive clumps were more likely to have a blue \(\delta v\) value, since the more massive clumps might be more likely to be gravitationally collapsing, but we observe no trend.

Another indicator of the potential for gravitational collapse is the virial parameter, \(\alpha_{vir}\). An \(\alpha_{vir} < \sim 2\) signifies that a clump is gravitational bound and may undergo collapse.  Many of our clumps are defined as gravitationally bound, but the \(\delta v\) value once again does not correlate with this parameter. It is important to note that in our calculation of \(\alpha_{vir}\) we only account for kinetic and potential energy but we do not account for external pressure or magnetic field support.  As discussed in \cite{Svoboda16}, a modest magnetic field of only a few hundred micro-Gauss is needed to raise the typical SCC virial parameter above 2.  A proper virial analysis also needs to consider the weight of the surrounding cloud on the clump and turbulent surface pressure (see \cite{Kirk17}). 

We were unable to find any significant trend between $\delta v$ and physical parameters of the clumps.
This statement includes the six best inflow candidates which are not strongly clustered in mass or
virial parameter.  Our lack of correlation is in contrast to the recent survey of 17 massive clumps that are $70$ $\mu$m dark by \citep{Traficante18} which did find a correlation between HCO$^+$ 1-0 line asymmetry (defined in a different method than used here) and the peak mass surface density of the clumps; however the peak surface densities in our survey are generally lower (see Figure \ref{sourceprop}) than the $0.1$ g cm$^{-3}$ threshold for which they observe more asymmetric line profiles.

\section{Analysis \& Discussion}

Only 6\% of observed starless clump candidates show a signature of inflow in HCO$^+$ 1-0.  
This result, taken without caveats, indicates that large-scale inflow motions are either a rare occurrence in our Galaxy or that a global initial inflow process during the starless phase is very quick, and is therefore only observable in a few percent of clumps.
\cite{Svoboda16} estimated the starless phase lifetime of clumps and showed that the phase lifetime varied inversely with the mass of the clumps.  If the low percentage of observable inflow signatures is attributable to a brief massive inflow phase, then the \citep{Svoboda16} relationship indicates that this massive global inflow starless phase lasts $\sim$ 60,000 years for the median inflow candidate in our sample.
However we must consider the caveats as there are many different factors that could result in non-detection of inflow. The detection of a blue asymmetric profiles requires a specific set of conditions; namely increasing excitation temperature along the line of sight, being optically thick along the line of sight, and a favorable line of sight velocity field. Simulations of global collapse show that there is a directional efficiency factor of up to $\sim 50$ ]\% for observing blue asymmetric line profiles with common dense gas tracers \citep{Smith2012, Smith2013}, meaning our observed inflow detection rate of 6\% could be 12\% in reality. Furthermore, we are only sensitive to motions that are fast enough to produce a measurable blue asymmetry meaning that a slow flow comparable to the FWHM linewidth observed in our large beam will not be detected.
Radiative transfer simulations \citep{Smith2013} also show that larger beam sizes suppress blue asymmetries.  
So, our observational result is likely a lower limit to the number of clumps with detectable inflow.
In this section, we shall model and analyze the profiles of the best inflow candidates, calculate the mass inflow rate, and the total change in mass of the clump in a free fall time. 

We model our best inflow candidates using the analytic HILL5 model from \cite{devries05} with 5-parameters: $\tau_0$ the optical 
depth at the peak observed line temperature  $T_{\rm pk}$, $v_{\rm in}$ the inflow velocity, $v_{\rm LSR}$, and $\sigma$ the
linewidth.  The HILL5 model solves the equation of radiative transfer with the assumption that
the excitation temperature is a linear function of optical depth (a "HILL" model).  \cite{devries05} found that the 5 parameter version of the HILL model most consistently and accurately recovered the inflow velocity when fit to synthetic spectra created from radiative transfer models compared to other parametrizations of the HILL model and the simpler ``two layer'' model of \cite{myers96}.  We apply Bayesian parameter estimation to fit the HILL5 model independently to each \hcop\  1-0 spectrum.  We use the affine-invariant Markov chain Monte Carlo (MCMC) ensemble sampler {\tt emcee} \citep{foremanmackey13} to maximize its log-likelihood function
\begin{equation}\label{eq:Likelihood}
\ln p(T|\boldsymbol{\theta}) = - \frac{1}{2} \sum_{\rm n} \left[ \frac{(T_{\rm n} - f_{\rm h5}(\boldsymbol{\theta}))^2}{\sigma_{T}^2} \right]
\end{equation}
over each channel $n$ in the spectrum with observed main beam temperature $T_{\rm n}$, uniform baseline RMS $\sigma_{T}$, and \hillf\ model value (Equation \ref{eq:Likelihood}) for parameter set $\boldsymbol{\theta}$.
Uniform priors were chosen with ranges between $\tau_0 = 0 - 10$, $\Delta v_{\rm LSR} = \pm 3$ \kms, $v_{\rm in} = 0 - 1$ km/s, $\sigma = 0 - 10$ \kms, and $T_{\rm pk} = 0 - 100$ K.  
This range was chosen to accommodate the range of physical values while preventing non-physical solutions when the model spectra extend outside the data range.
Initial guesses start with $\tau_0 = 1$, $\Delta v_{\rm LSR} = 0$ \kms, $v_{\rm in} = 0$ km/s, $\sigma = \sigma_{\rm{NH}_3}$ \kms, and $T_{\rm pk} = 2T_{\rm{HCO}^+}$ K.

\begin{figure}
\centering
\includegraphics[scale=.6]{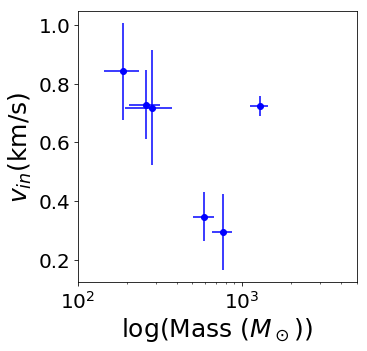}
\caption{ Here we show the relationship between infall velocity as modeled by HILL5 and clump mass. We do not find any significant correlation here. The Spearmann rank correlation coefficient is -0.66. We are working with a low number of objects, and our \(\delta v\) analysis showed that there is no correlation between radius (a factor in \(v_{in}\)) and inflow likelihood as represented by \(\delta v\). (See figure \ref{4panel}). Infall velocity and other HILL5 results can be found in the Appendix   }
\label{mcmc}
\end{figure}

The modeling results from the clump BGPS 3302 are shown in Figure \ref{mcmc}.  The inflow velocity is well constrained to be $v_{in} = 0.72 \pm 0.03$ km/s.  We find good fits to all of the best inflow candidates with inflow velocities that range from $0.25$ km/s to $0.84$ km/s.  The median inflow velocity is $0.72$ km/s and the median uncertainty from the MCMC analysis is $0.13$ km/s.  These motions, if attributed to inflow, are supersonic. We also modeled two additional sources (BGPS 2533 and BGPS 3151) which had nearly equal temperature blue and red peaks.  Unsurprisingly, the HILL5 model is consistent with zero inflow velocity in these two cases. 

The inflow velocities are plotted versus mass of the clumps in the upper left panel of Figure \ref{4panel}. The inflow velocities do not correlate with the mass of the clumps; however, the inflow candidates with masses below the median 300 M$_{\odot}$ of the sample all tend to have the largest inflow velocities.

Using the inflow velocity estimated by our HILL5 modeling, we calculate the mass inflow rate taken from \cite{Lopezsepulcre10}:

\begin{eqnarray}
\dot{M} & = & 4\pi R^2 v_{in} \rho = \frac{3 M v_{in}}{R}  \nonumber \\
        & = & 3068 \frac{\rm{M}_{\odot}}{\rm{Myr}} \bigg(\frac{M}{1000 \rm{M}_{\odot}}\bigg)\bigg(\frac{v_{in}}{1 km/s}\bigg)\bigg(\frac{1 pc}{R}\bigg) \;\;.
\end{eqnarray}
The highest inflow rate is BGPS 3302 with 1880 M$_{\odot}$/Myr and the lowest inflow rate is BGPS 4029 with  520 M$_{\odot}$/Myr.  
We are assuming complete symmetrical collapse of the clump with all of the mass of the clump participating in the inflow motions. The calculated mass inflow rate will be an overestimate if only a fraction of the mass of the clump is associated with inflow motions within the 12m telescope beam. On the other hand, we only measure velocities along the line of sight, and this geometry likely underestimates the inflow velocity and subsequent mass accretion rate.  It is not clear which factor would dominate and therefore our mass accretion rates are highly uncertain. 

Large scale inflow has been observed in both high-mass and low-mass star-forming regions. \cite{Peretto13} studied star forming regions centered at a massive inflowing filament, similar to the type of object we are searching for. They found a mass inflow rate of $2500 \pm 1000$ M$_{\odot}$/Myr, which overlaps with the upper range of \(\dot{M}\) seen in our sample; however, this clump is known to already be forming massive stars, and is not starless. 
\cite{Liu17} observed large-scale inflows near 1000 M$_{\odot}$/Myr from maps of HCO$^+$ 1-0 toward three protostellar BGPS clumps that were also observed in the MALT90 survey. \cite{Barnes10} observed a dense molecular core undergoing gravitational collapse and also used Hill5 modeling and HCO$^+$ observations. They found a mass inflow rate of \(3400 \pm 1700 M_{\odot}/Myr\) which is a higher rate than most of our candidates, but does overlap with our highest inflow rates. \cite{Kirk13} observed a filament inflowing into a central region in the intermediate-mass Serpens star forming cloud and found that along the filament $\dot{M}=30$ M$_{\odot}$/Myr and radially the rate was $130$ M$_{\odot}$/Myr. These values are smaller than the range in which we observe, however the object they were observing was lower in mass than our inflow candidates ($\sim 20$ M$_{\odot}$ in the filament). 
It is doubtful that we could detect such low inflow rates toward BGPS SCCs given the modest linewidths and our requirement that we observe a clear self-absorption dip in the HCO$^+$ 1-0 line profile.  Setting a lower limit of $v_{in} = 0.2$ km/s to produce a self-absorption dip for the typical BGPS SCC linewidth, then for a 0.5 pc, 100 M$_{\odot}$ clump, we have a limiting detectable mass inflow rate of $\sim 120$ M$_{\odot}$/Myr.  Thus, it is entirely possible that we have missed the types of inflow deteced in the Serpens star-forming region and are only detecting massive, large-scale flows. 

The candidate inflow clumps in our survey appear to be in an earlier phase than the examples where large scale flows are observed that are described above.  
The only published inflow survey of comparable objects with calculated mass inflow rates is the study of \cite{He2016} which analyzed spectra towards pre-stellar clumps observed in the ATLASGAL and MALT90 surveys.  The objects are classified as pre-stellar based on the lack of mid-infrared or far-infrared compact source emission.  \cite{He2016} find a median mass inflow rate of 2600 M$_{\odot}$/Myr which is higher than our mass inflow estimates and comparable to the mass inflow rate observed by \cite{Peretto13} toward  a massive star-forming hub-filament complex. 
The \cite{He2016} inflow sample is generally more massive ($500$ M$_{\odot}$) and has higher peak mass surface densities ($0.14$ g cm$^{-3}$) by a factor of $3-4$ than our sample.  As noted in \S3.1, our constraints to qualify a source as an inflow candidate are more stringent than in the \cite{He2016}.
    
For our six best inflow candidates, 
we calculate the change in mass of the clump assuming a constant mass accretion rate over the average clump free fall time ($\Delta M = \dot{M} t_{ff}$).  Using the median free-fall time for each clump calculated in \cite{Svoboda16}, we find that 
BGPS 4029 has the lowest \(\Delta M\) of \(310 M_{\odot}\) and BGPS 2432 has the highest value of \(1410 M_{\odot}\). 
If we divide the total change in mass during the free fall time by the mass of the clump we find that most of the clumps would double in mass over a free fall time (See Table 2).
It should be noted that the free fall time corresponds to the time for the clump to collapse at the average clump volume density and corresponds to a median value of $0.7$ Myr for the inflow candidates.  
This timescale is longer than the median gas depletion timescale of $M/\dot{M} = 0.5$ Myr for the inflow candidates which may indicate that the clumps are growing in mass from the surrounding medium. 
While most SCCs may have a phase lifetime longer than their clump-average free fall timescale \citep{Svoboda16}, the inflow candidates may be rare examples of clumps that are in a phase where they are embedded within a supersonic flow and are growing in mass at a significant rate.

Our single-dish survey, while a good first step for identifying promising inflow candidates, cannot exclude the possibility that observed line profiles are due to unresolved motions between dense cores within the clump.  Furthermore, our analysis of the mass inflow rates have made the very simple assumption that the total mass of the clump is associated with the inflow motions.
These objects are clearly interesting targets for higher resolution observations (i.e. with \textit{ALMA}) which can directly address these caveats.  

\section{Conclusions}
We observed 101 massive Starless Clump Candidates from the BGPS survey using the optically thick, intermediate density gas tracer HCO$^+$ J=1-0.
We found only 6 clumps (6\%) had line profiles that had a self-absorption dip at nearly the same velocity (slightly redshifted) as the peak of NH$_3$ (1,1) emission and had a HCO$^+$ 1-0 peak intensity blue-ward of its self-absorption dip (blue asymmetry). 
The complete sample of SCCs has a small blue excess of only \(E=0.03\) indicating that blue asymmetric line profiles are rare toward BGPS starless clump candidates in HCO$^+$ 1-0.  
We found no correlation between $\delta v$ and the physical parameters of the clumps such as mass and virial parameter. 
Using HILL5 radiative transfer modeling of the best inflow candidates we determine inflow velocities and mass inflow rates that range from 500 - 2000 M$_{\odot}$/Myr.  At these accretion rates, the SCCs would double in mass over a free-fall time.
These clumps may be in a phase where they are embedded within a supersonic inflow and are growing in mass at a significant rate.

 The 12m ARO telescope has a beam size that is about the same size or slightly larger than the clump radii we were observing.  As a result, our 6 inflow candidates are only just candidates for inflow motions.  Higher spatial resolution mapping observations on telescopes such as the \textit{GBT} and \textit{ALMA} are necessary to confirm whether these objects are truly starless and whether they have global inflow motions.

\section{Acknowledgments}\label{sec:Acknowledgements}
We sincerely thank the staff and operators of the Arizona Radio Observatory.
JKC, BES, and YLS are supported in part by NSF grant AST-1410190.
BES is also supported by the NSF Graduate Research Fellowship under grant No.~DGE-114395.  JKC thanks the Arizona NASA Space Grant Consortium for internship support.

\facilities{
    ARO:12m
}

\software{\tt
    \href{https://ds9.si.edu}{ds9},
    \href{https://www.iram.fr/IRAMFR/GILDAS/}{GILDAS CLASS}
}

\appendix

The Appendix contains Table 1 listing the 100 objects targeted using HCO$^+$ with their distance (RA, Dec), distance in kpc, peak \(v_{LSR}\), \(\sigma_{v_{LSR}}\), \(\sigma_{\delta v}\) value, and \(\sigma_{\delta v}\). Figures 1 and 2 contain the 96 spectra of HCO$^+$ detections with their NH$_3$ peak indicated by a red dashed line. Figure 3 has the HILL5 results of the best six candidates. The results are shown in a triangle graph showing the parameters in columns (from left to right) and rows (top to bottom) corresponding to optical depth, \(v_{LSR}\), infall velocity, dispersion, and peak \(T_{mb}\).

\begin{deluxetable}{cccccccc}
    \tablecolumns{7} 
    \tablewidth{0pt}
    \tablecaption{Starless Clump Candidates}
     \tablehead{
     Source &
     \(\alpha\) & 
     \(\delta\) & 
     Distance & 
     \(v_{LSR}\)(HCO$^+$)$_{pk}$ &
     \(\sigma_{v_{LSR}}\) &
     \(\delta v\) &
     \(\sigma_{\delta v}\) \\
     &
     (hh:mm:ss)\tablenotemark{a} &
     (dd:mm:ss) & (kpc) & (km/s) & (km/s)  & &}
    \startdata
BGPS 2427	&	18	:	9	:	33.88	&	-20	:	47	:	0.76	&	4.670	&	30.491	&	0.045	&	-0.108	&	0.032	\\
BGPS 2430	&	18	:	8	:	49.41	&	-20	:	40	:	23.82	&	5.013	&	22.262	&	0.094	&	0.353	&	0.035	\\
BGPS 2432	&	18	:	9	:	44.59	&	-20	:	47	:	10.21	&	4.369	&	29.002	&	0.036	&	-1.236	&	0.036	\\
BGPS 2437	&	18	:	10	:	19.41	&	-20	:	50	:	27.45	&	4.437	&	-1.464	&	0.01	&	0.305	&	0.018	\\
BGPS 2533	&	18	:	10	:	30.29	&	-20	:	14	:	44.2	&	4.975	&	30.623	&	0.121	&	-1.247	&	0.129	\\
BGPS 2564	&	18	:	10	:	6.08	&	-18	:	46	:	5.64	&	3.013	&	29.404	&	0.038	&	-0.263	&	0.069	\\
BGPS 2693	&	18	:	11	:	13.56	&	-17	:	44	:	54.85	&	2.103	&	18.695	&	0.115	&	-0.874	&	0.230	\\
BGPS 2710	&	18	:	13	:	49.04	&	-17	:	59	:	33.25	&	1.200	&	34.246	&	0.027	&	-0.354	&	0.029	\\
BGPS 2724	&	18	:	14	:	13.61	&	-17	:	59	:	52.02	&	1.185	&	34.428	&	0.115	&	-0.543	&	0.045	\\
BGPS 2732	&	18	:	14	:	26.85	&	-17	:	58	:	50.93	&	1.191	&	35.082	&	0.07	&	-1.189	&	0.084	\\
BGPS 2742	&	18	:	14	:	29.1	&	-17	:	57	:	21.83	&	1.183	&	34.528	&	0.057	&	-0.542	&	0.038	\\
BGPS 2762	&	18	:	11	:	39.52	&	-17	:	32	:	9.4	&	3.304	&	18.597	&	0.062	&	0.459	&	0.046	\\
BGPS 2931	&	18	:	17	:	27.51	&	-17	:	6	:	8.42	&	3.285	&	23.119	&	0.04	&	0.564	&	0.067	\\
BGPS 2940	&	18	:	17	:	17.15	&	-17	:	1	:	7.47	&	3.366	&	19.903	&	0.013	&	-0.061	&	0.009	\\
BGPS 2945	&	18	:	17	:	27.35	&	-17	:	0	:	23.66	&	1.178	&	23.031	&	0.013	&	0.585	&	0.027	\\
BGPS 2949	&	18	:	17	:	33.74	&	-16	:	59	:	34.94	&	1.279	&	22.352	&	0.025	&	-0.217	&	0.040	\\
BGPS 2970	&	18	:	17	:	5.08	&	-16	:	43	:	28.66	&	3.568	&	40.159	&	0.02	&	0.102	&	0.016	\\
BGPS 2971	&	18	:	16	:	48.12	&	-16	:	41	:	8.91	&	1.815	&	40.371	&	0.059	&	3.434	&	0.160	\\
BGPS 2976	&	18	:	17	:	7.84	&	-16	:	41	:	14.59	&	1.830	&	40.413	&	0.022	&	0.902	&	0.032	\\
BGPS 2984	&	18	:	18	:	18.23	&	-16	:	44	:	52.26	&	1.855	&	17.818	&	0.026	&	-1.170	&	0.062	\\
BGPS 2986	&	18	:	18	:	29.68	&	-16	:	44	:	50.69	&	2.014	&	20.446	&	0.017	&	0.660	&	0.025	\\
BGPS 3018	&	18	:	19	:	13.88	&	-16	:	35	:	16.47	&	6.522	&	19.27	&	0.045	&	0.281	&	0.043	\\
BGPS 3030	&	18	:	19	:	19.68	&	-16	:	31	:	39.82	&	1.784	&	18.887	&	0.027	&	-0.180	&	0.046	\\
BGPS 3110	&	18	:	20	:	16.27	&	-16	:	8	:	51.13	&	2.005	&	17.217	&	0.006	&	-0.219	&	0.008	\\
BGPS 3114	&	18	:	20	:	31.5	&	-16	:	8	:	37.8	&	1.848	&	23.525	&	0.016	&	0.229	&	0.005	\\
BGPS 3117	&	18	:	20	:	6.68	&	-16	:	4	:	45.75	&	2.001	&	18.519	&	0.022	&	-0.032	&	0.028	\\
BGPS 3118	&	18	:	20	:	16.17	&	-16	:	5	:	50.72	&	2.001	&	16.214	&	0.017	&	-0.557	&	0.030	\\
BGPS 3125	&	18	:	20	:	6.11	&	-16	:	1	:	58.02	&	3.466	&	18.139	&	0.046	&	-4.215	&	0.216	\\
BGPS 3128	&	18	:	20	:	35.27	&	-16	:	4	:	53.81	&	4.170	&	19.044	&	0.011	&	-0.396	&	0.029	\\
BGPS 3129	&	18	:	20	:	12.99	&	-16	:	0	:	24.13	&	4.289	&	18.812	&	0.046	&	-0.781	&	0.080	\\
BGPS 3134	&	18	:	19	:	52.72	&	-15	:	56	:	1.56	&	4.083	&	20.658	&	0.004	&	0.139	&	0.010	\\
BGPS 3139	&	18	:	20	:	34.24	&	-15	:	58	:	14	&	5.408	&	21.103	&	0.037	&	-0.805	&	0.057	\\
BGPS 3151	&	18	:	20	:	23.19	&	-15	:	39	:	31.96	&	3.379	&	38.86	&	0.041	&	-0.614	&	0.048	\\
BGPS 3220	&	18	:	24	:	57.03	&	-13	:	20	:	32.39	&	3.874	&	46.916	&	0.024	&	0.190	&	0.009	\\
BGPS 3243	&	18	:	25	:	32.74	&	-13	:	1	:	31.05	&	4.597	&	68.069	&	0.019	&	-0.547	&	0.030	\\
BGPS 3247	&	18	:	25	:	14.45	&	-12	:	54	:	16.74	&	4.447	&	44.886	&	0.2	&	-0.344	&	0.244	\\
BGPS 3276	&	18	:	26	:	24.92	&	-12	:	49	:	30.07	&	3.379	&	66.515	&	0.047	&	-0.521	&	0.046	\\
BGPS 3300	&	18	:	26	:	28.42	&	-12	:	37	:	3.98	&	11.668	&	62.427	&	0.031	&	-0.704	&	0.031	\\
BGPS 3302	&	18	:	27	:	15.23	&	-12	:	42	:	56.45	&	11.785	&	64.66	&	0.044	&	-0.967	&	0.033	\\
BGPS 3306	&	18	:	23	:	34.02	&	-12	:	13	:	52.79	&	4.777	&	57.082	&	0.016	&	-0.030	&	0.023	\\
BGPS 3312	&	18	:	25	:	44.52	&	-12	:	28	:	34.11	&	5.271	&	47.098	&	0.012	&	-0.292	&	0.100	\\
BGPS 3315	&	18	:	25	:	33.24	&	-12	:	26	:	50.63	&	4.793	&	47.678	&	0.021	&	2.111	&	0.184	\\
BGPS 3344	&	18	:	26	:	40	&	-12	:	25	:	15.81	&	4.314	&	65.583	&	0.19	&	-0.132	&	0.227	\\
BGPS 3442	&	18	:	28	:	13.51	&	-11	:	40	:	44.94	&	3.442	&	67.007	&	0.003	&	1.165	&	0.022	\\
BGPS 3444	&	18	:	28	:	27.26	&	-11	:	41	:	33.99	&	3.297	&	69.749	&	0.082	&	0.061	&	0.081	\\
BGPS 3475	&	18	:	28	:	28.28	&	-11	:	6	:	44.16	&	3.426	&	77.643	&	0.27	&	0.641	&	0.108	\\
BGPS 3484	&	18	:	29	:	15.74	&	-10	:	58	:	28.73	&	3.484	&	56.316	&	0.023	&	-0.017	&	0.041	\\
BGPS 3487	&	18	:	29	:	22.77	&	-10	:	58	:	1.69	&	3.490	&	53.283	&	0.061	&	-0.459	&	0.073	\\
BGPS 3534	&	18	:	30	:	33.45	&	-10	:	24	:	19	&	3.225	&	\nodata	&	\nodata	&	\nodata	&	\nodata	\\
BGPS 3604	&	18	:	30	:	43.92	&	-9	:	34	:	42.15	&	11.010	&	50.725	&	0.013	&	-0.770	&	0.017	\\
BGPS 3606	&	18	:	29	:	41.95	&	-9	:	24	:	49.1	&	4.214	&	49.143	&	0.028	&	-0.419	&	0.054	\\
BGPS 3608	&	18	:	31	:	54.82	&	-9	:	39	:	5.03	&	4.081	&	61.869	&	0.121	&	-0.597	&	0.041	\\
BGPS 3627	&	18	:	31	:	42.32	&	-9	:	24	:	29.17	&	4.169	&	82.71	&	0.126	&	1.181	&	0.115	\\
BGPS 3656	&	18	:	32	:	49.54	&	-9	:	21	:	29.26	&	3.906	&	77.655	&	0.061	&	0.215	&	0.042	\\
BGPS 3686	&	18	:	34	:	14.58	&	-9	:	18	:	35.84	&	2.939	&	76.927	&	0.121	&	-0.301	&	0.103	\\
BGPS 3705	&	18	:	34	:	32.69	&	-9	:	14	:	9.4	&	3.116	&	63.207	&	0.013	&	0.771	&	0.048	\\
BGPS 3710	&	18	:	34	:	20.55	&	-9	:	10	:	1.94	&	2.505	&	74.405	&	0.024	&	-0.113	&	0.013	\\
BGPS 3716	&	18	:	34	:	24.15	&	-9	:	8	:	3.6	&	3.146	&	75.039	&	0.082	&	-0.225	&	0.027	\\
BGPS 3736	&	18	:	33	:	28.22	&	-8	:	55	:	4.36	&	5.031	&	65.62	&	0.016	&	0.139	&	0.018	\\
BGPS 3822	&	18	:	33	:	32.06	&	-8	:	32	:	26.27	&	3.370	&	55.14	&	0.045	&	0.389	&	0.035	\\
BGPS 3833	&	18	:	33	:	36.5	&	-8	:	30	:	50.7	&	4.493	&	56.113	&	0.044	&	0.403	&	0.036	\\
BGPS 3892	&	18	:	35	:	59.74	&	-8	:	38	:	56.48	&	5.300	&	62.643	&	0.027	&	-0.816	&	0.015	\\
BGPS 3922	&	18	:	33	:	40.98	&	-8	:	14	:	55.3	&	9.876	&	89.299	&	0.021	&	0.071	&	0.022	\\
BGPS 3924	&	18	:	34	:	51.17	&	-8	:	23	:	40.02	&	5.782	&	81.11	&	0.029	&	-0.089	&	0.017	\\
BGPS 3982	&	18	:	34	:	30.79	&	-8	:	2	:	7.36	&	11.582	&	54.019	&	0.054	&	0.109	&	0.073	\\
BGPS 4029	&	18	:	35	:	54.4	&	-7	:	59	:	44.6	&	3.539	&	80.687	&	0.027	&	-0.458	&	0.022	\\
BGPS 4082	&	18	:	35	:	10.07	&	-7	:	39	:	43.55	&	5.084	&	98.994	&	0.027	&	-0.433	&	0.024	\\
BGPS 4085	&	18	:	33	:	57.05	&	-7	:	29	:	31.43	&	5.087	&	96.409	&	0.025	&	-0.131	&	0.026	\\
BGPS 4095	&	18	:	35	:	4	&	-7	:	36	:	6.46	&	5.353	&	114.384	&	0.06	&	1.359	&	0.061	\\
BGPS 4119	&	18	:	36	:	29.65	&	-7	:	42	:	6.09	&	5.577	&	54.339	&	0.136	&	-0.352	&	0.050	\\
BGPS 4135	&	18	:	37	:	44.06	&	-7	:	48	:	15.35	&	3.572	&	62.11	&	0.047	&	0.618	&	0.022	\\
BGPS 4140	&	18	:	36	:	49.66	&	-7	:	40	:	36.83	&	3.617	&	96.106	&	0.025	&	0.114	&	0.032	\\
BGPS 4145	&	18	:	36	:	52.95	&	-7	:	39	:	49.2	&	4.985	&	96.52	&	0.023	&	-0.010	&	0.022	\\
BGPS 4191	&	18	:	37	:	4.58	&	-7	:	33	:	12.26	&	5.123	&	97.211	&	0.019	&	-0.226	&	0.016	\\
BGPS 4230	&	18	:	35	:	50.85	&	-7	:	12	:	23.58	&	5.034	&	109.337	&	0.052	&	1.604	&	0.065	\\
BGPS 4294	&	18	:	38	:	51.58	&	-6	:	55	:	36.52	&	5.688	&	58.936	&	0.146	&	2.106	&	0.115	\\
BGPS 4297	&	18	:	38	:	56.37	&	-6	:	55	:	8.44	&	4.973	&	59.254	&	0.025	&	0.464	&	0.024	\\
BGPS 4346	&	18	:	38	:	49.58	&	-6	:	31	:	27.06	&	5.823	&	95.56	&	0.111	&	0.741	&	0.030	\\
BGPS 4347	&	18	:	38	:	42.93	&	-6	:	30	:	27.83	&	5.295	&	95.789	&	0.066	&	1.196	&	0.037	\\
BGPS 4354	&	18	:	38	:	51.42	&	-6	:	29	:	15.38	&	5.840	&	93.949	&	0.085	&	-0.013	&	0.045	\\
BGPS 4356	&	18	:	37	:	29.48	&	-6	:	18	:	12.13	&	4.453	&	110.176	&	0.021	&	0.143	&	0.014	\\
BGPS 4375	&	18	:	39	:	10.19	&	-6	:	21	:	15.9	&	3.793	&	92.913	&	0.014	&	-0.160	&	0.019	\\
BGPS 4396	&	18	:	38	:	34.74	&	-5	:	56	:	43.97	&	4.266	&	112.998	&	0.027	&	0.241	&	0.026	\\
BGPS 4402	&	18	:	39	:	28.64	&	-5	:	57	:	58.57	&	4.285	&	100.149	&	0.124	&	0.679	&	0.091	\\
BGPS 4422	&	18	:	38	:	47.88	&	-5	:	36	:	16.38	&	3.917	&	111.503	&	0.049	&	0.728	&	0.054	\\
BGPS 4472	&	18	:	41	:	17.32	&	-5	:	9	:	56.83	&	3.216	&	47.937	&	0.084	&	1.138	&	0.093	\\
BGPS 4732	&	18	:	44	:	23.4	&	-4	:	2	:	1.21	&	3.782	&	89.609	&	0.074	&	1.020	&	0.060	\\
BGPS 4827	&	18	:	44	:	42.45	&	-3	:	44	:	21.63	&	4.928	&	89.611	&	0.04	&	3.011	&	0.082	\\
BGPS 4841	&	18	:	42	:	15.65	&	-3	:	22	:	26.19	&	4.266	&	83.534	&	0.031	&	-0.288	&	0.022	\\
BGPS 4902	&	18	:	46	:	11.36	&	-3	:	42	:	55.73	&	4.656	&	84.262	&	0.136	&	0.050	&	0.056	\\
BGPS 4953	&	18	:	45	:	51.82	&	-3	:	26	:	24.16	&	5.502	&	90.192	&	0.054	&	-0.328	&	0.033	\\
BGPS 4962	&	18	:	45	:	59.61	&	-3	:	25	:	14.53	&	6.092	&	88.765	&	0.078	&	0.493	&	0.091	\\
BGPS 4967	&	18	:	43	:	27.8	&	-3	:	5	:	14.94	&	3.681	&	81.2	&	0.021	&	0.697	&	0.025	\\
BGPS 5021	&	18	:	44	:	37.07	&	-2	:	55	:	4.4	&	5.181	&	78.77	&	0.072	&	-0.843	&	0.049	\\
BGPS 5064	&	18	:	45	:	48.44	&	-2	:	44	:	31.65	&	5.210	&	100.608	&	0.009	&	-0.081	&	0.011	\\
BGPS 5089	&	18	:	48	:	49.88	&	-2	:	59	:	47.86	&	6.534	&	85.137	&	0.012	&	-0.042	&	0.015	\\
BGPS 5090	&	18	:	46	:	35.81	&	-2	:	42	:	30.19	&	5.181	&	95.464	&	0.04	&	-0.554	&	0.036	\\
BGPS 5114	&	18	:	50	:	23.54	&	-3	:	1	:	31.58	&	3.681	&	67.254	&	0.02	&	0.802	&	0.015	\\
BGPS 5166	&	18	:	47	:	54.26	&	-2	:	26	:	7.11	&	6.092	&	102.403	&	0.019	&	-0.263	&	0.024	\\
BGPS 5183	&	18	:	47	:	0.29	&	-2	:	16	:	38.63	&	6.534	&	\nodata	&	\nodata	&	\nodata	&	\nodata	\\
BGPS 5243	&	18	:	47	:	54.7	&	-2	:	11	:	10.72	&	5.210	&	96.62	&	0.048	&	0.708	&	0.048	\\
\enddata
\tablenotetext{a}{All coordinates are epoch J2000.0.}
\tablenotetext{b}{Distances are the maximum likelihood distance of the DPDF (Distance Probability Density Function) in \cite{Svoboda16}}
\end{deluxetable}

\begin{deluxetable}{ccccccc}
\tablecolumns{8} 
\tablewidth{0pt}
\tablecaption{Inflow Candidates}
\label{bestinflowtable}
\tablehead{Source	&	\(v_{in}\)	&	\(\sigma_{v in}\)	&	\(t_{ff}\)\tablenotemark{a}	&	\(\dot{M}\)		&	\(\Delta M\)	&	\(\Delta M/M\) \\
   &  (km/s) & (km/s) & (Myr) & (M$_{\odot}$/Myr) & (M$_{\odot}$) & 
}
\startdata
BGPS 2432	&	0.718	&	0.196	&	0.661	&	823	&	544	&	2.9	\\
BGPS 3300	&	0.841	&	0.165	&	0.871	&	607	&	529	&	3.2	\\
BGPS 3302	&	0.723	&	0.034	&	0.751	&	1876	&	1408	&	1.4	\\
BGPS 3604	&	0.728	&	0.119	&	0.704	&	748	&	526	&	2.8	\\
BGPS 4029	&	0.296	&	0.130	&	0.593	&	521	&	309	&	0.7	\\
BGPS 5021	&	0.347	&	0.082	&	0.694	&	567	&	393	&	1.0 \\
\enddata
\tablenotetext{a}{The median of the free-fall probability density function calculated in \citep{Svoboda16}.}
\end{deluxetable}

\begin{figure}
    \centering
    \includegraphics[scale=.5]{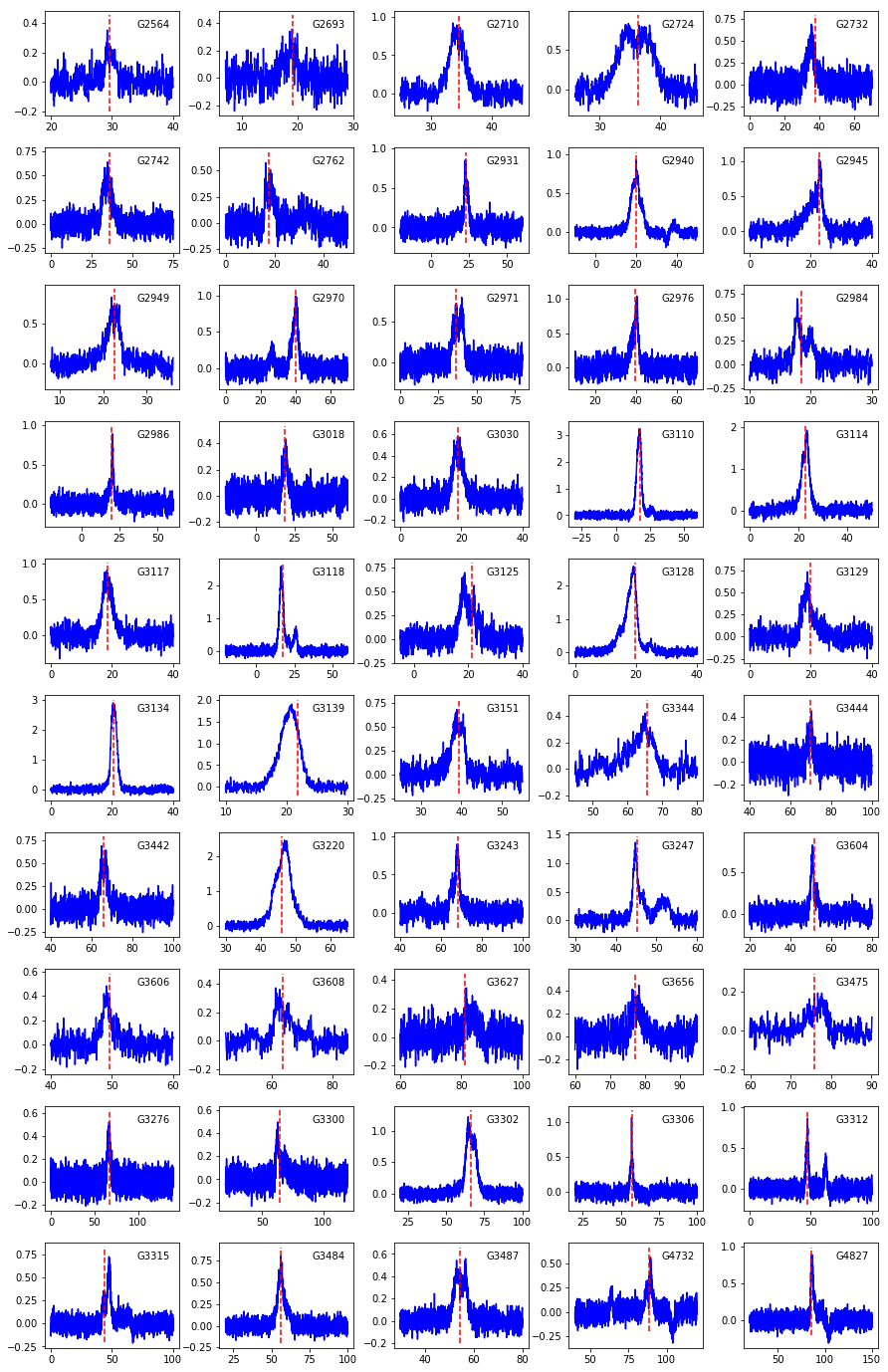}
    \caption{The line profiles of the first half of the 99 clumps that have detections in HCO$^+$. The y-axis is in km/s and the x-axis is in Kelvin. The red dashed line shows the observed ammonia peak.}
    \label{allspectra1}
\end{figure}

\begin{figure}
    \centering
    \includegraphics[scale=.42]{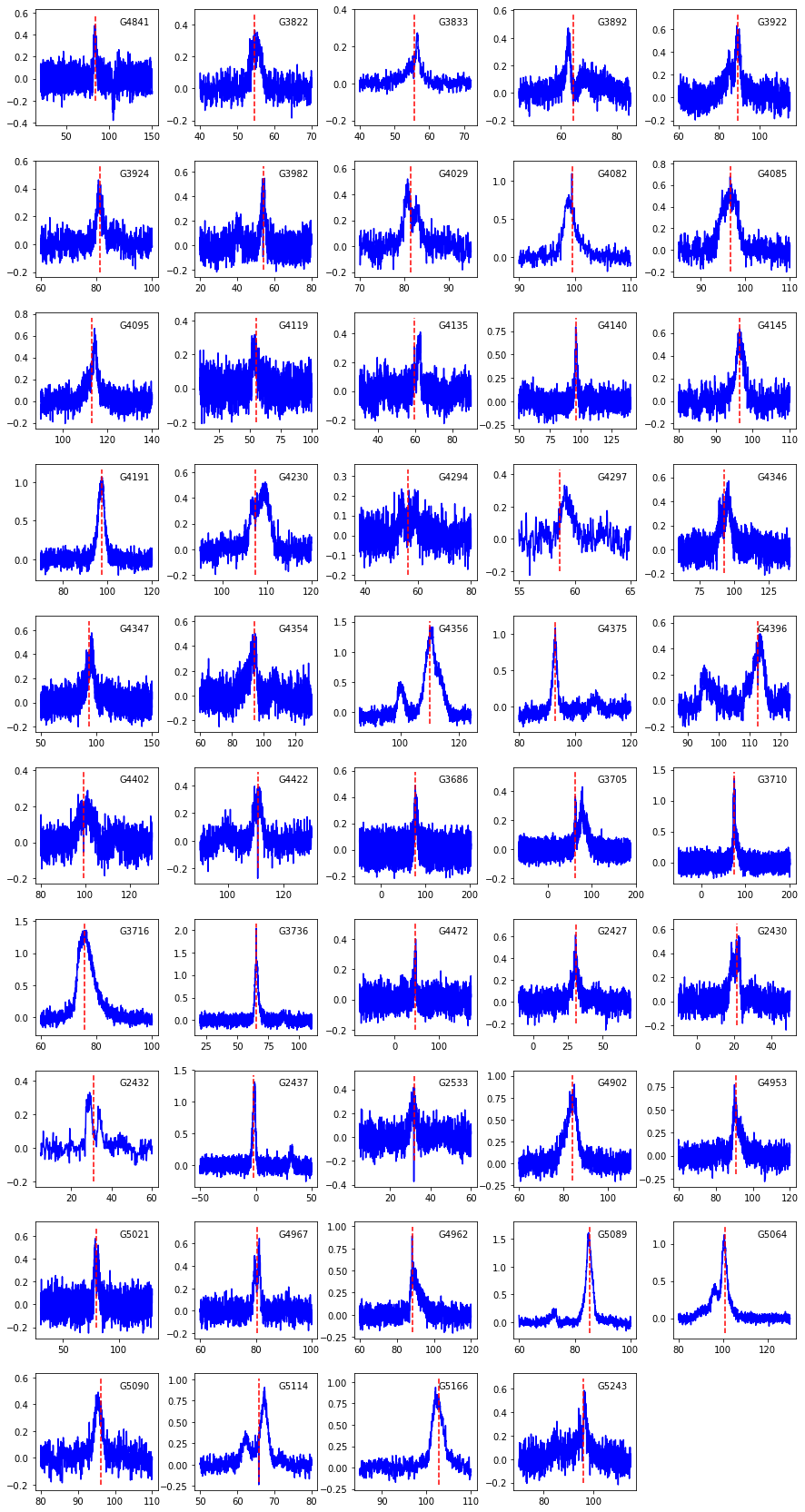}
    \caption{The line profiles of the second half of the 99 clumps that have detections in HCO$^+$. The y-axis is in km/s and the x-axis is in Kelvin.The red dashed line shows the observed ammonia peak.}
    \label{allspectra2}
\end{figure}

\begin{figure}
    \centering
    \includegraphics[scale=1]{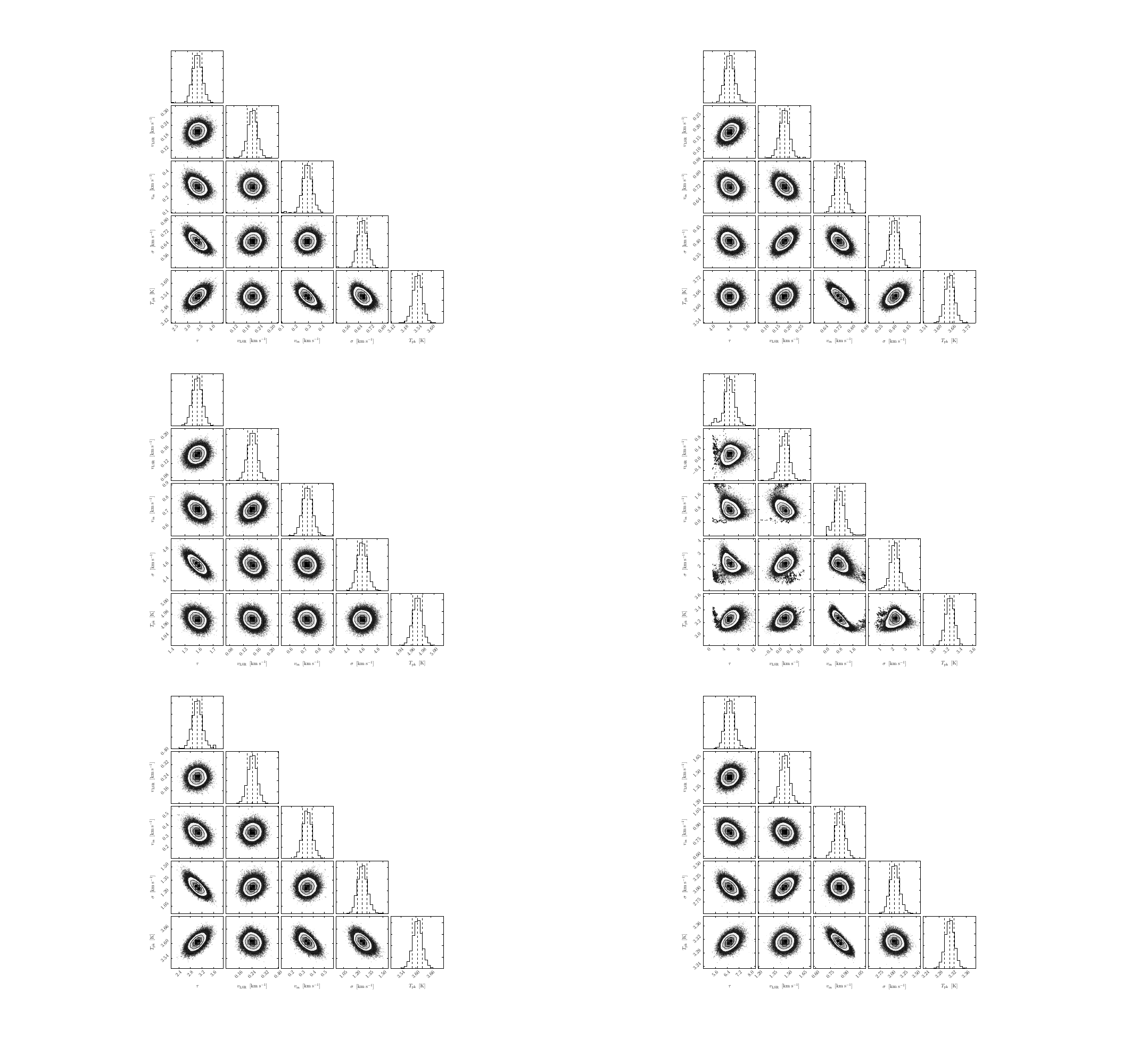}
    \caption{Triangle plots of the results of the five HILL5 modeling parameters (\(\tau\), \(v_{LSR}\), \(v_{in}\), \(\sigma\), and \(T_{pk}\)) for the best infall candidates. Shown starting from the upper left are 4029, 3604, 3302, 2432, 5021, and 3300 ending at the lower right.}
    \label{fig:my_label}
\end{figure}




\clearpage

\end{document}